\documentclass[conference]{IEEEtran}
\usepackage{graphicx} 
\usepackage{cite}
\usepackage{comment}
\usepackage{amsmath,amssymb,amsfonts}
\usepackage{booktabs}
\usepackage{wrapfig}
\usepackage{float}
\usepackage{parcolumns}
\usepackage{array}
\usepackage{multicol}
\usepackage[normalem]{ulem}
\usepackage{xcolor}

\usepackage{calc}

\newlength{\Acolwidth}
\newlength{\Bcolwidth}
\newcommand{\Abox}[1]{\makebox[\Acolwidth][c]{$#1$}}
\newcommand{\Bbox}[1]{\makebox[\Bcolwidth][c]{$#1$}}

\begin{document}
\title{

A Wide-Regulation-Range Hybrid \\ Switched-Capacitor  Converter for 48V Automotive Power Systems

}

\author{%
  \IEEEauthorblockN{Georgios Spanodimos}
  \IEEEauthorblockA{%
    Electrical and Computer Engineering\\
    University of California, Los Angeles\\
    Los Angeles, USA\\
    georspano22@ucla.edu
  }

\and

  \IEEEauthorblockN{Guanyu Qian}
  \IEEEauthorblockA{%
    Electrical and Computer Engineering\\
    University of California, Los Angeles\\
    Los Angeles, USA\\
    gyqian@ucla.edu
  }

\and

  \IEEEauthorblockN{Xiaofan Cui}
  \IEEEauthorblockA{%
    Electrical and Computer Engineering\\
    University of California, Los Angeles\\
    Los Angeles, USA\\
    cuixf@seas.ucla.edu
  }

}

\maketitle

\begin{abstract}
This paper presents a hybrid switched-capacitor converter (HSCC) with a novel multi-mode modulation (3M) scheme for wide-range voltage regulation in 48-V automotive power systems. By introducing a three-state operating sequence beyond the conventional 2:1 resonant operation, the proposed converter achieves variable step-down conversion ratios while preserving soft-switching operation in most transitions. The proposed modulation supports both zero-current switching (ZCS) and zero-voltage switching (ZVS) modes, enabling efficient operation over a broad range of load and conversion conditions. To enable voltage regulation, a closed-loop control configuration is proposed with a linear proportional-integral (PI) controller, with gain tuning assisted by reinforcement learning (RL) to address the converter's nonlinear and variable-frequency nature while maintaining good transient performance. A hardware prototype was built to validate the proposed modulation scheme. The measured results verify ZCS operation over voltage conversion ratios of 0.2--0.4, with a peak efficiency exceeding 92\% at 100~W, and efficiency above 88\% over a wide operating range for 3:1 conversion. The feasibility of both ZCS and ZVS operation is also experimentally demonstrated. These results show that the proposed HSCC significantly extends the practical regulation range of resonant switched-capacitor converters while maintaining high efficiency.
\end{abstract}

\section{Introduction}

The role of modern automobiles is expanding beyond transportation to serve as real-time edge artificial intelligence (AI) nodes. With the deployment of autonomous driving systems, vehicles are increasingly required to execute complex AI workloads locally, resulting in a substantial rise in onboard computing power demands from hundreds of watts to several kilowatts in next-generation vehicles\cite{Kotb2023PowerTrends, DeOliveira2020AApplications}. The conventional 12V lead-acid battery-based architecture is no longer sufficient to support such high-performance computing and electrified loads.
The 48V low-voltage bus has emerged as a promising solution, offering a 4x reduction in current and a 16x reduction in power delivery network (PDN) conduction losses \cite{Wood2025High-VoltageNetworks}. However, this transition introduces new challenges for the power conversion systems (PCS) that interface the 48V bus with diverse legacy and emerging automotive loads. These PCS must accommodate a wide range of voltage levels, from 48V power steering and pumps \cite{Kotb2023PowerTrends} to traditional 12V subsystems to sub-5V point-of-load (PoL) supplies for onboard XPUs, while meeting stringent requirements on efficiency, power density, and transient performance.

In this paper, we propose a hybrid switched-capacitor (HSCC) voltage regulator with a new multi-mode modulation (3M) scheme, a potential next-generation PCS for 48V automotive architectures. The proposed PCS is highly adaptable to different low-voltage system configurations. It can operate either as a front-end converter from 48V to 12V to support legacy subsystems, or as a local PoL converter directly stepping down from 48V to sub-5V levels for computing loads. High efficiency is maintained across this wide conversion ratio and operating range.

\begin{figure}[t]
    \centering
    \includegraphics[width=0.96\linewidth]{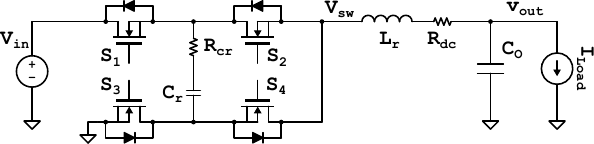}
    \caption{Schematic of the hybrid switched-capacitor converter.}
    \label{fig:CKT}
\end{figure}

The proposed PCS adopts a flying-capacitor multi-level (FCML) switched-capacitor topology with a small output resonant inductor. Compared with conventional buck-based solutions, this approach significantly reduces the output inductor size, thereby achieving higher power density. In addition, the converter features tight voltage regulation, wide step-down conversion ratios, and high efficiency, which are rarely achieved simultaneously in existing switched-capacitor-based designs.
Conventional automotive systems often employ DC--DC converters based on transformer-isolated topologies~\cite{Bellur2007DC-DCApplications}. These converters, such as the dual active bridge (DAB) and the LLC resonant converter, transfer energy across a magnetic core with primary and secondary windings. The transformer turn ratio provides the voltage scaling needed to achieve high conversion ratios. However, the transformer contributes significant weight and volume, and its non-ideal magnetic characteristics, such as leakage inductance and core losses, become problematic at high switching frequencies, posing substantial challenges to improving power density. This is particularly important in EV systems, where reducing overall system weight directly benefits vehicle efficiency and range. Switched-capacitor converters (SCCs) have emerged as a promising alternative to magnetics-based converters, offering improved power density by replacing bulky magnetic components with capacitors as the primary energy-transfer element. Traditional SCCs, however, are constrained by an inherently fixed and efficient 2:1 conversion ratio, and their efficiency degrades sharply when operated away from this natural ratio due to hard-switching losses. An HSCC, shown in Fig.~\ref{fig:CKT}, which incorporates a small output inductor to form an $LC$ resonant tank, is designed to eliminate hard-switching losses through soft-switching operation. Recent advancements in multi-resonant operation~\cite{Xie2025PerformanceRegulation, Cervera2015ARatio, Turhan2015Step-downRatio, Jong2019ResonantFrequencies} have addressed the fixed-ratio limitation by introducing multiple resonant states and idle intervals within each switching cycle, demonstrating that these converters can achieve high efficiency across both light- and heavy-load conditions. Still, most studies have centered on operation at the natural 2:1 conversion ratio~\cite{Chen2020Two-StageBus},\cite{Sambo2024AConverters} since this point enables straightforward open-loop soft-switching conditions. However, the variable-conversion-ratio modulation in this resonant topology, without sacrificing efficiency, remains largely unexplored. Moreover, closed-loop implementations capable of tightly regulating the output voltage have received little attention.

This paper also introduces a new closed-loop control configuration that integrates the proposed three-mode modulation (3M) with reinforcement learning (RL) for tuning controller parameters. Specifically, the proposed scheme combines an additional buck-like discharging state with series-resonant operation, while maintaining ZCS and ZVS for most switching events, thereby improving efficiency across varying load conditions. An adaptive zero-crossing detection algorithm is developed to automatically identify ZCS conditions under parameter variations and operating uncertainties, such that the buck-mode interval and resonant-period modulation can be adjusted to achieve wide-range and fast voltage regulation, unlike prior approaches that rely on idle intervals and frequency modulation~\cite{Xie2025PerformanceRegulation}. Given the highly nonlinear dynamics and multi-objective nature of the proposed PCS, an offline RL agent is incorporated to tune the control parameters and achieve near-optimal transient performance without relying on an accurate system model. The experimental results validate that the proposed converter achieves a continuously variable conversion ratio from 2:1 to 5:1 while maintaining voltage regulation, with peak efficiency exceeding 92\% at 100~W. In particular, at a 3:1 conversion ratio, efficiency stays above 88\% under all-load operation.

\section{Theory of Operation}

\subsection{Converter Operation and Dynamics}

The proposed modulation scheme differs from the traditional two-state fixed-duty-ratio operation, which is typically limited to a voltage conversion ratio of approximately half the input voltage. In contrast, the proposed scheme operates in three states.  The timing notation of each state, together with its physical interpretation, is defined as follows:
\begin{itemize}
    \item \textbf{State 1} (\(T_1:[t_0,t_1]\)): \(C_r\) is charged from the input.
    \item \textbf{State 2} (\(T_2:[t_1,t_2]\)): \(C_{\text{out}}\) discharges to ground.
    \item \textbf{State 3} (\(T_3:[t_2,t_3]\)): \(C_r\) delivers charge to \(C_{\text{out}}\).
\end{itemize}

\begin{figure*}[t]
    \centering
    \includegraphics[width=0.96\linewidth]{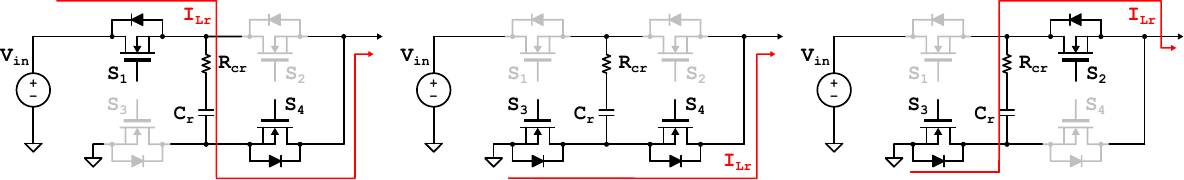}
    \caption{inductor current commutation paths in States~1--3, from left to right.}
    \label{fig:commutation}
\end{figure*}

In the proposed modulation scheme, the converter transitions from State~1 to State~3. The corresponding inductor current path for each state is illustrated in Fig.~\ref{fig:commutation}. The introduction of State~2 provides a faster discharge path for the voltage across \(C_r\), thereby allowing \(V_{cr}\) to deviate from the fixed one-half-input-voltage in the steady state. As a result, a variable voltage conversion ratio can be achieved.

To facilitate the analysis of the converter dynamics, we formulate the system as a switched affine system of the form

\begin{equation}
\dot{x}=A_{\sigma}x+b_{\sigma},
\qquad \sigma\in\{1,2,3\},
\end{equation}
with the state vector defined as
\begin{equation}
x=
\begin{bmatrix}
i_L & v_{cr} & v_{out}
\end{bmatrix}^{T}.
\end{equation}
Hence, the state-space matrices for the three operating states can be derived as
\begin{equation*}
\scalebox{0.85}{$
\begin{aligned}
A_1 &=
\begin{bmatrix}
\Abox{-\dfrac{R_{\mathrm{eq1}}}{L_r}} & \Abox{-\dfrac{1}{L_r}} & \Abox{-\dfrac{1}{L_r}}\\[8pt]
\Abox{\dfrac{1}{C_r}} & \Abox{0} & \Abox{0}\\[8pt]
\Abox{\dfrac{1}{C_o}} & \Abox{0} & \Abox{0}
\end{bmatrix},
&\quad
B_1 &=
\begin{bmatrix}
\Bbox{\dfrac{V_{\mathrm{in}}}{L_r}} & \Bbox{0}\\[8pt]
\Bbox{0} & \Bbox{0}\\[8pt]
\Bbox{0} & \Bbox{-\dfrac{I_{\mathrm{Load}}}{C_o}}
\end{bmatrix},
\\[12pt]
A_2 &=
\begin{bmatrix}
\Abox{-\dfrac{R_{\mathrm{eq2}}}{L_r}} & \Abox{0} & \Abox{-\dfrac{1}{L_r}}\\[8pt]
\Abox{0} & \Abox{0} & \Abox{0}\\[8pt]
\Abox{\dfrac{1}{C_o}} & \Abox{0} & \Abox{0}
\end{bmatrix},
&\quad
B_2 &=
\begin{bmatrix}
\Bbox{0} & \Bbox{0}\\[8pt]
\Bbox{0} & \Bbox{0}\\[8pt]
\Bbox{0} & \Bbox{-\dfrac{I_{\mathrm{Load}}}{C_o}}
\end{bmatrix},
\\[12pt]
A_3 &=
\begin{bmatrix}
\Abox{-\dfrac{R_{\mathrm{eq3}}}{L_r}} & \Abox{\dfrac{1}{L_r}} & \Abox{-\dfrac{1}{L_r}}\\[8pt]
\Abox{-\dfrac{1}{C_r}} & \Abox{0} & \Abox{0}\\[8pt]
\Abox{\dfrac{1}{C_o}} & \Abox{0} & \Abox{0}
\end{bmatrix},
&\quad
B_3 &=
\begin{bmatrix}
\Bbox{0} & \Bbox{0}\\[8pt]
\Bbox{0} & \Bbox{0}\\[8pt]
\Bbox{0} & \Bbox{-\dfrac{I_{\mathrm{Load}}}{C_o}}
\end{bmatrix}.
\end{aligned}
$}
\end{equation*}

The equivalent lumped parasitic resistances are defined as follows to simplify the notation:
\begin{equation}\label{eq:Req}
\begin{aligned}
R_{\mathrm{eq1}} = R_{\mathrm{eq3}} &= 2R_{\mathrm{DS,ON}} + R_{dc} + R_{cr}, \\
R_{\mathrm{eq2}} &= 2R_{\mathrm{DS,ON}} + R_{dc},
\end{aligned}
\end{equation}
where $R_{\mathrm{DS,ON}}$ denotes the on-resistance of the switch, $R_{dc}$ is the dc resistance of the inductor, and $R_{cr}$ is the equivalent series resistance (ESR) of the flying capacitor. As a result, the state trajectory over one complete switching period is governed by the following evolution: 
\begin{equation}{\label{eq:sol1}}
\begin{aligned}
x(T)
&= \Phi_3 \Phi_2 \Phi_1 x(0)
   + \Phi_3 \Phi_2 \Gamma_1
   + \Phi_3 \Gamma_2
   + \Gamma_3 ,
\end{aligned}
\end{equation}
where the corresponding time evolution operator $\Phi_\sigma$ and zero-state response term $\Gamma_\sigma$ are defined as
\begin{equation}{\label{eq:sol2}}
\Phi_\sigma = e^{A_\sigma T_\sigma},
\qquad
\Gamma_\sigma = \int_0^{T_\sigma} e^{A_\sigma(T_\sigma-\tau)} B_\sigma\, d\tau,
\end{equation}

For the variable voltage-conversion range, \(T_1\) can be treated as an independently controlled variable, while \(T_2\) and \(T_3\) are determined by the corresponding zero-current or zero-voltage instants.

\subsection{ZCS Mode}
In the proposed ZCS operation, two switching transitions are maintained under ZCS conditions: the transitions of switches \(S_3\) and \(S_4\) from State~2 to State~3, and of \(S_2\) and \(S_3\) from State~3 to State~1. The following initial conditions are defined:
\begin{equation}
\label{eq:IC_il}
i_{L_r}(t_0) = i_{L_r}(t_2) = i_{L_r}(t_3) = 0~\mathrm{A}
\end{equation}
\begin{equation}
\label{eq:IC_vcr}
v_{C_r}(t_0) = V_{C_r,\min}, \qquad
v_{C_r}(t_1) = V_{C_r,\max}.
\end{equation}
Based on the state evolution in \eqref{eq:sol1}--\eqref{eq:sol2}, and under the small-ripple assumption that \(v_{\mathrm{out}} \approx V_{\mathrm{out}}\) remains approximately constant over all three states, closed-form expressions for the inductor current trajectories in the three operating states are obtained as follows.

In State~1, the inductor current \(i_{L_r}\) is driven by the \(L_r\)--\(C_r\) resonant tank and charges \(C_r\). The natural frequency, damping coefficient, and damped resonant frequency are defined as
\begin{equation}\label{eq:freq_def}
\omega_o = \frac{1}{\sqrt{L_r C_r}}, \qquad
a = \frac{R_{\mathrm{eq1}}}{2L_r}, \qquad
\omega_d = \sqrt{\omega_o^2-a^2}.
\end{equation}
Then, under the initial conditions in \eqref{eq:IC_il} and \eqref{eq:IC_vcr}, the inductor current in State~1 is given by
\begin{equation}\label{eq:state1}
i_{L_r,1}(t)
=
A e^{-a(t-t_0)}\sin\!\big(\omega_d (t-t_0)\big),
\end{equation}
with
\begin{equation}\label{eq:state1_param}
A = \frac{V_{\mathrm{in}} - V_{C_r,\min} - V_{\mathrm{out}}}{\omega_d L_r}.
\end{equation}

In State~2, \(C_r\) is floating, and its voltage therefore does not affect the inductor current trajectory during this interval. Moreover, the inductor current in State~2 follows the same form as that of a conventional buck converter during the off-state. For notational convenience, define
\begin{equation}\label{eq:state2_param}
B = i_{L_r,1}(t_1), \qquad
b = \frac{R_{\mathrm{eq2}}}{L_r}.
\end{equation}
Then, the inductor current in State~2 can be expressed as
\begin{equation}\label{eq:state2}
i_{L_r,2}(t)
=
\left(B+\frac{V_{\mathrm{out}}}{R_{\mathrm{eq2}}}\right)e^{-b(t-t_1)}
-\frac{V_{\mathrm{out}}}{R_{\mathrm{eq2}}}.
\end{equation}

By solving the ZCS condition in~\eqref{eq:IC_il}, the duration, \(T_2\), from \(t_1\) to \(t_2\) can be obtained as
\begin{equation}\label{eq:state2_zcs}
T_2 = t_2-t_1
=
\frac{L_r}{R_{\mathrm{eq2}}}
\ln\!\left(
\frac{B R_{\mathrm{eq2}}}{V_{\mathrm{out}}}+1
\right).
\end{equation}
In practice, this interval is not imposed by an external clock, but is determined by the zero-current detection instant. 

\begin{figure*}[t]
    \centering
    \includegraphics[width=0.48\textwidth]{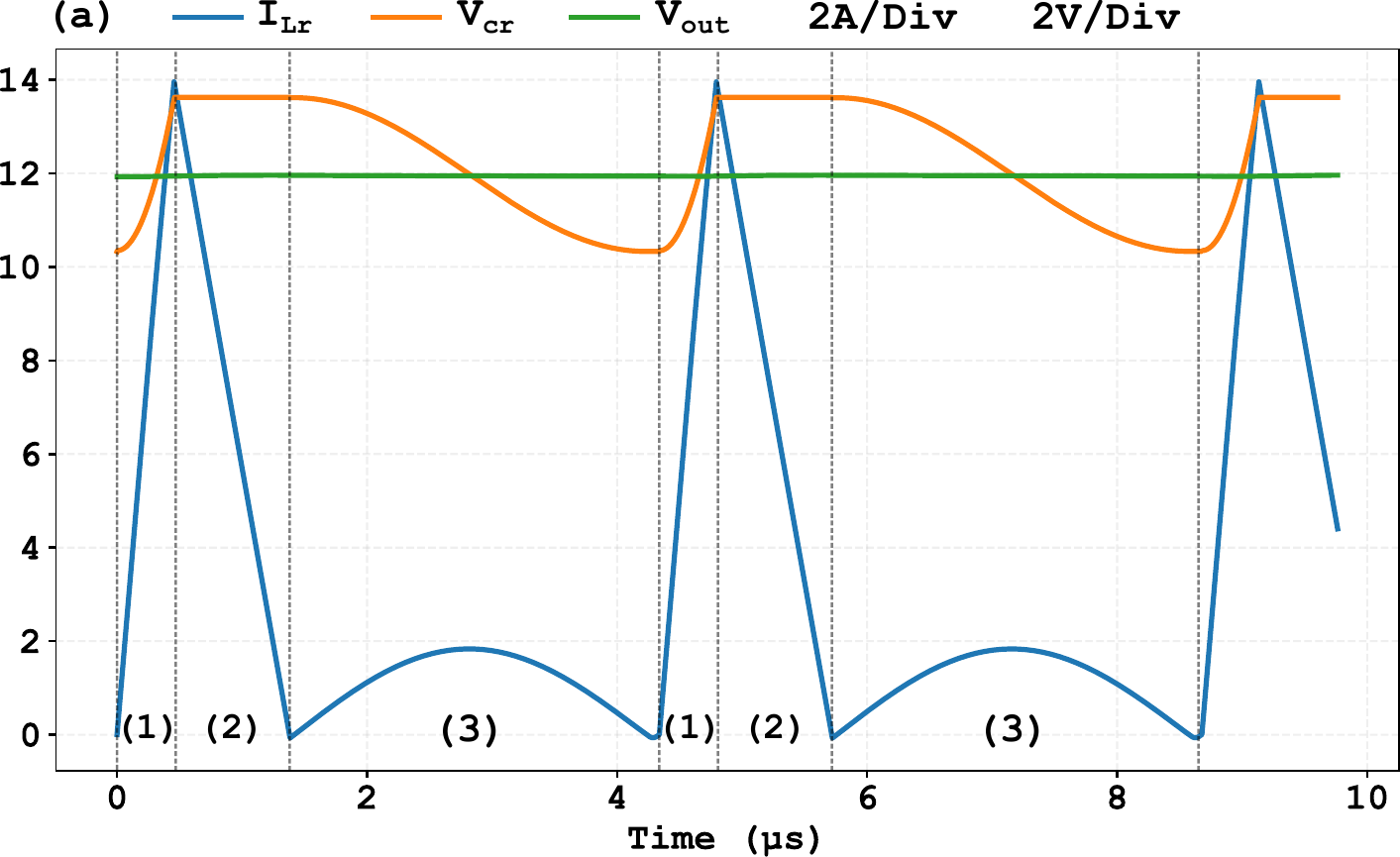}
    \hfill
    \includegraphics[width=0.48\textwidth]{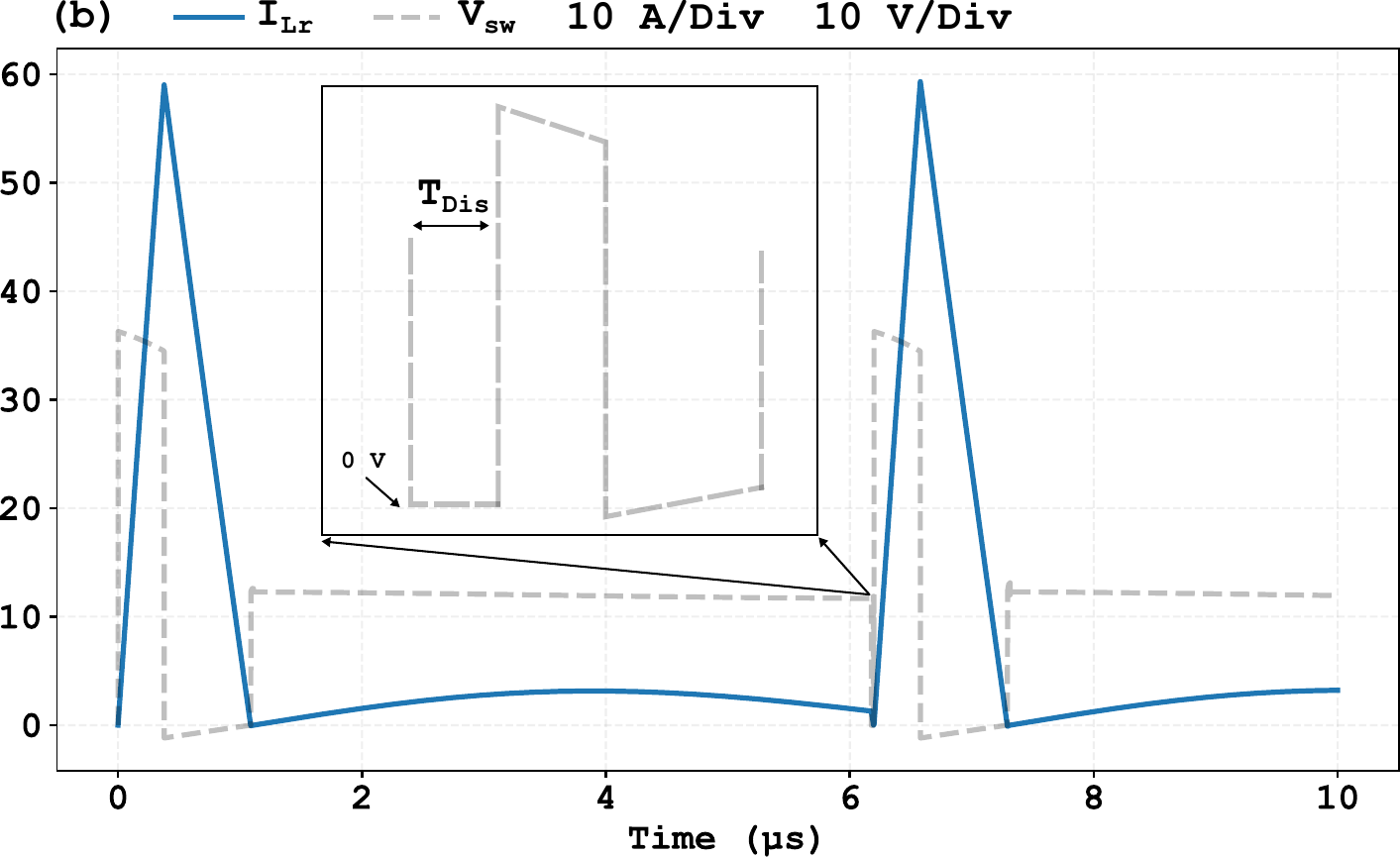}
    \caption{(a) ZCS-mode waveforms for 48-V-to-12-V operation, showing \(v_{\mathrm{out}}\), \(v_{C_r}\), and \(i_{Lr}\) with the corresponding states indicated. The transitions from State~1 to State~2 and from State~2 to State~3 are under ZCS. (b) ZVS-mode waveforms, where the transition from State~3 to State~1 achieves ZVS. The switching-node voltage is also labeled.}
    \label{fig:waveform}
\end{figure*}

State~3 follows the same resonant form as State~1, but with different initial conditions. Accordingly, the inductor current in this interval can be written as
\begin{equation}\label{eq:state3}
i_{L_r,3}(t)
=
C e^{-a(t-t_2)}\sin\!\big(\omega_d (t-t_2)\big),
\end{equation}
where
\begin{equation}\label{eq:state3_param}
C = \frac{V_{C_r,\max} - V_{\mathrm{out}}}{\omega_d L_r}.
\end{equation}
The duration of State~3 is determined by the ZCS condition and is equal to one-half of the damped resonant period:
\begin{equation}\label{eq:state3_zcs}
T_3 = t_3-t_2 = \frac{\pi}{\omega_d},
\end{equation}
which can likewise be determined by the zero-current sensing scheme. A representative waveform for the 48-V-to-12-V operation is shown in Fig.~\ref{fig:waveform}, where \(i_L\), together with \(v_{\mathrm{out}}\) and \(v_{C_r}\), is labeled for each state.

Furthermore, with the definitions of the auxiliary variables \(\alpha\) and \(\beta\),
\begin{align}
\alpha &= 1 - e^{-aT_1}\left(\frac{a}{\omega_d}\sin(\omega_d T_1)+\cos(\omega_d T_1)\right) \\[8pt]
\beta &= 1 + e^{-a\pi/\omega_d},
\end{align}
the analytical expressions for \(V_{C_r,\min}\) and \(V_{C_r,\max}\) can be determined as
\begin{align}
V_{C_r,\text{min}} \label{eq:vcrmin_expr}
&= \frac{\alpha(1-\beta)(V_{\text{in}}-V_{\text{out}})+\beta V_{\text{out}}}
{\beta+\alpha(1-\beta)} \\[12pt]
V_{C_r,\text{max}} \label{eq:vcrmax_expr}
&= V_{C_r,\text{min}} + \left(V_{\text{in}}-V_{C_r,\text{min}}-V_{\text{out}}\right)\alpha.
\end{align}

Hence, by substituting \eqref{eq:vcrmin_expr} and \eqref{eq:vcrmax_expr} into \eqref{eq:state1}--\eqref{eq:state3}, the inductor current trajectories can be obtained in closed form as functions of \(L_r\), \(C_r\), \(T_1\), \(V_{\text{in}}\), and \(V_{\text{out}}\).

\subsection{ZVS Mode}
To eliminate turn-on switching loss by recovering the energy stored in the transistor parasitic output capacitances \(C_{\mathrm{OSS}}\), ZVS operation is desirable. In addition to ZCS mode, the proposed modulation scheme also supports a ZVS mode by introducing an additional discharge interval \(T_{\mathrm{dis}}\). Although the proposed modulation scheme is capable of achieving ZVS for both the high-side and low-side switches~\cite{Sambo2024AConverters}, this work focuses on a mixed operating mode that combines ZCS and ZVS transitions, namely, ZCS for the State~2 to State~3 transition and ZVS for the State~3 to State~1 transition. For ease of reference, this operating mode is referred to as the ZVS mode throughout the following sections. Therefore, only the case in which the low-side switch \(S_4\) achieves ZVS is presented in Fig.~\ref{fig:waveform}(b).

In the ZVS mode,  to fully discharge the \(C_{\mathrm{OSS}}\) of \(S_4\), the converter enters an additional interval \(T_{\mathrm{dis}}\), during which the required switches remain non-conducting and \(S_3\) is turned on to provide a return path for discharging the parasitic output capacitance of \(S_4\), thereby ensuring ZVS turn-on. The determination of the duration \(T_{\mathrm{dis}}\) therefore requires careful consideration. At the beginning of the ZVS transition, the output capacitance \(C_{\mathrm{OSS}}\) of \(S_4\) is charged to the nominal drain-to-source voltage \(V_{\mathrm{DS}}\). To achieve ZVS turn-on, the inductor current must provide sufficient energy to discharge the \(C_{\mathrm{OSS}}\) of \(S_4\) from \(V_{\mathrm{DS}}\) to zero. Therefore, the inductor current at the end of State~3 must remain positive, and we define the current as \(I_{\mathrm{OFF}}\) and initiate the resonant discharge of \(C_{\mathrm{OSS}}\).

To determine the required magnitude of \(I_{\mathrm{OFF}}\), the charge stored in \(C_{\mathrm{OSS}}\) must first be quantified. This requires characterizing the transistor output capacitance, which commonly exhibits a nonlinear dependence on the drain-to-source voltage~\cite{Kasper2016ZVSRevisited,Ye2019ADensity}. The equivalent charge-based capacitance \(C_{Q,\mathrm{eq}}(V_{\mathrm{DS}})\) is defined as
\begin{equation}\label{eq:ZVS}
C_{Q,\mathrm{eq}}(V_{\mathrm{DS}})
=
\frac{Q_{\mathrm{OSS}}(V_{\mathrm{DS}})}{V_{\mathrm{DS}}}
=
\frac{1}{V_{\mathrm{DS}}}\int_{0}^{V_{\mathrm{DS}}} C_{\mathrm{OSS}}(v)\,dv
\end{equation}
where \(Q_{\mathrm{OSS}}(V_{\mathrm{DS}})\) is the total charge stored in \(C_{\mathrm{OSS}}\) at the given \(V_{\mathrm{DS}}\). This definition captures the effective capacitance over the full charging range.

To achieve full discharge, the energy stored in \(C_{\mathrm{OSS}}\) at a given \(V_{\mathrm{DS}}\) must not exceed the energy available in the inductor \(L_r\) at the start of the transition, which gives
\begin{equation}
\label{ZVS2}
\frac{1}{2}L_r I_{\mathrm{OFF}}^2
\ge
C_{Q,\mathrm{eq}}(V_{\mathrm{DS}})\,V_{\mathrm{DS}}^2.
\end{equation}
Accordingly, the minimum required value of \(I_{\mathrm{OFF}}\) is
\begin{equation}
\label{eq:ZVS3}
I_{\mathrm{OFF}}
\ge
\sqrt{\frac{2\,C_{Q,\mathrm{eq}}(V_{\mathrm{DS}})\,V_{\mathrm{DS}}^2}{L_r}}.
\end{equation}

For the ZVS transition of \(S_4\), \(C_{\mathrm{OSS}}\) is initially charged to \(V_{\mathrm{out}}\), i.e., \(V_{\mathrm{DS}}=V_{\mathrm{out}}\). By substituting \eqref{eq:ZVS3} into \eqref{eq:state3}, an analytical boundary for \(T_{\mathrm{dis}}\) can be obtained. However, accurate characterization of \(C_{Q,\mathrm{eq}}\) and an explicit analytical solution for \(T_{\mathrm{dis}}\) may require considerable effort. Hence, a practical approximation for the discharge interval is adopted from~\cite{Sambo2024AConverters,Ye2019ADensity}:
\begin{equation}
    \label{ZVS4}
    T_{\mathrm{dis}} \approx \pi \sqrt{\frac{L_r C_{\mathrm{OSS}}}{2}}.
\end{equation}

An identical analytical procedure can also be applied to the ZVS transition of the high-side switch \(S_1\).

\begin{figure*}[t]
    \centering
    \includegraphics[width=0.96\linewidth]{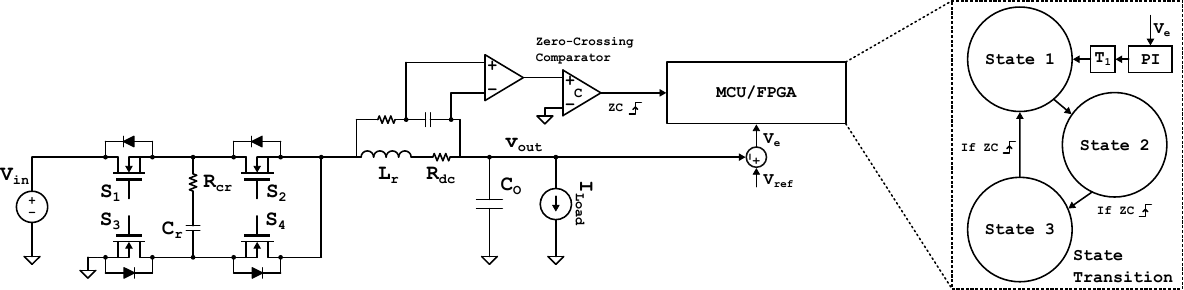}
    \caption{Proposed modulation with closed-loop control implementation and the corresponding state logic.}
    \label{fig:impl}
\end{figure*}

\begin{figure}[t]
    \centering
    \includegraphics[width=0.96\linewidth]{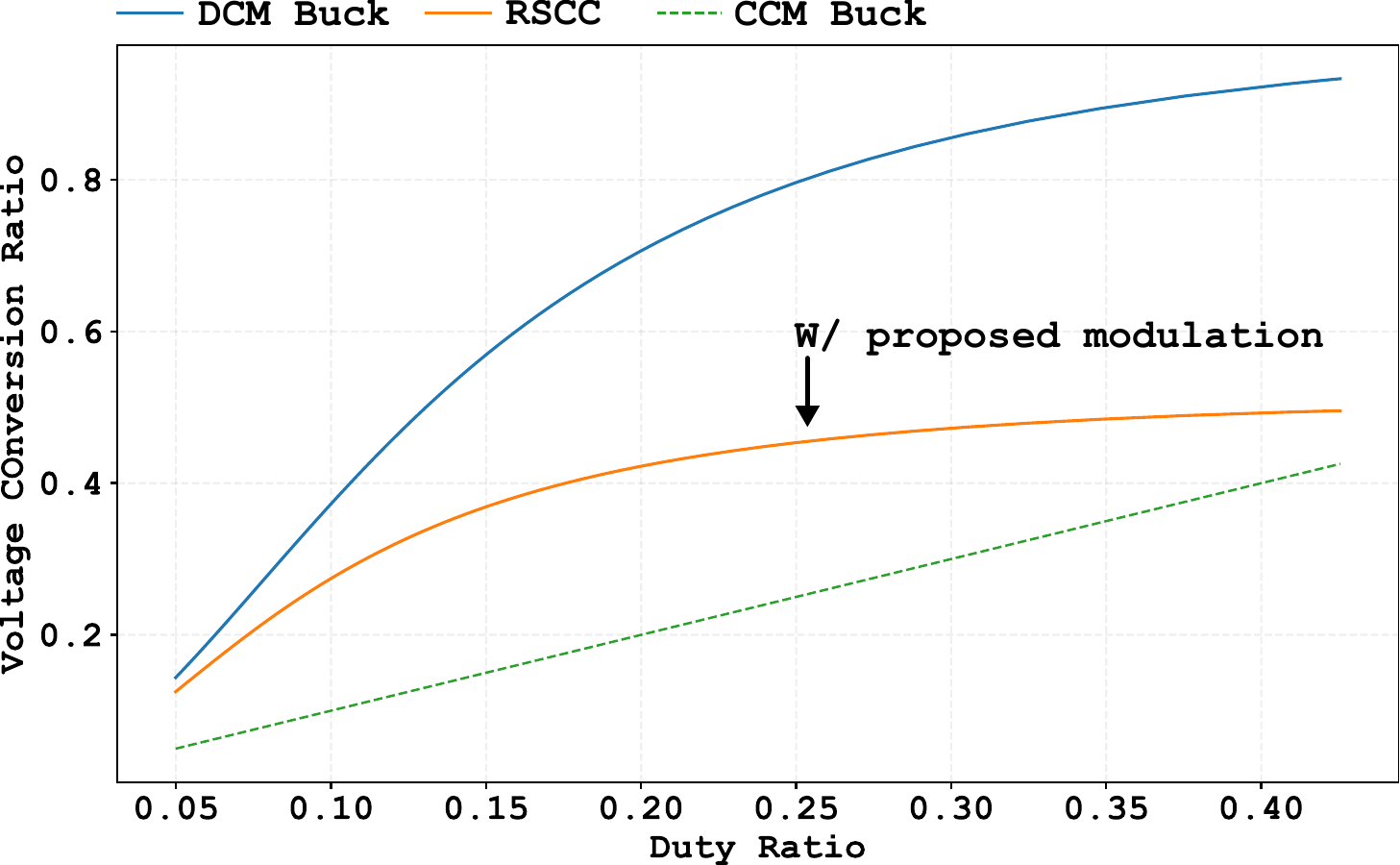}
    \caption{Comparison of voltage conversion ratios at \(V_{\mathrm{in}}=48~\mathrm{V}\) and \(I_{\mathrm{load}}=20~\mathrm{A}\) for the DCM buck converter, the HSCC with the proposed modulation, and the CCM buck converter using the passive component values listed in Table~I.}
    \label{fig:conv_ratio}
\end{figure}

\subsection{Closed-Loop Control for Variable Conversion Ratio}
The proposed modulation scheme can be directly applied to closed-loop voltage regulation. In this section, the focus is placed on the ZCS mode, and the detailed circuit implementation and state logic are shown in Fig.~\ref{fig:impl}. Under this modulation scheme, the voltage conversion ratio can be derived in a manner similar to that of a conventional discontinuous-conduction-mode (DCM) buck converter. Define the switching period as \(T=T_1+T_2+T_3\) and the duty ratio as \(D=T_1/T\). By enforcing the condition \(\langle i_L \rangle = \langle i_{\mathrm{Load}} \rangle\) and substituting \eqref{eq:state1}--\eqref{eq:state3}, a closed-form analytical expression for the voltage conversion ratio can be derived as a function of \(D\), \(L_r\), \(C_r\), \(R_{\mathrm{eq}}\), \(V_{\mathrm{in}}\), and \(I_{\mathrm{load}}\).

\begin{figure}[t]
    \centering
    \includegraphics[width=0.96\linewidth]{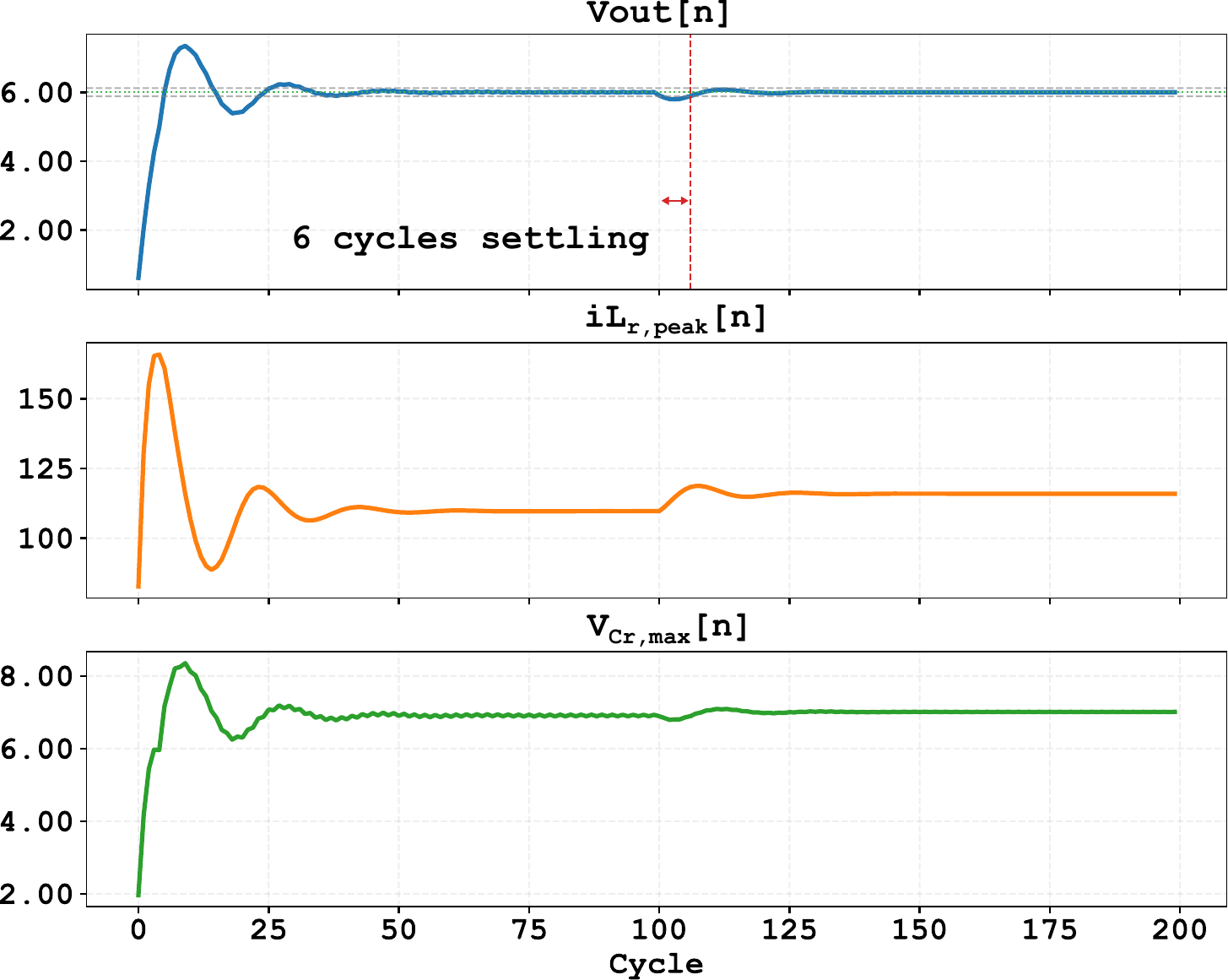}
    \caption{Simulated time-domain waveforms of \(v_{\mathrm{out}}\), \(i_{L_r}\), and \(v_{C_r}\) for 48-V-to-6-V conversion at \(I_{\mathrm{load}}=20~\mathrm{A}\)  under a 10\% load transient.}
    \label{fig:trans}
\end{figure}

Although the analytical expression is omitted here for brevity, a numerical comparison of the conversion ratio among the conventional DCM buck converter, the HSCC under the proposed modulation, and the continuous-conduction-mode (CCM) buck converter is shown in Fig.~\ref{fig:conv_ratio}, based on the passive-component values listed in Table~I and the operating point \(V_{\mathrm{in}}=48~\mathrm{V}\) and \(I_{\mathrm{load}}=20~\mathrm{A}\). As \(D\) approaches 0.5, the operation converges to that of a conventional 2:1 resonant converter, beyond which the conversion ratio can no longer be further increased while maintaining the proper ZCS condition. Nevertheless, the proposed high-step-down modulation is primarily intended for operation in the region \(D<0.5\). Furthermore, the proposed scheme is inherently a variable-frequency modulation. The variable \(T_1\) can be generated from the voltage error by a proportional-integral (PI) controller, whereas \(T_2\) and \(T_3\) are determined by the state evolution and zero-crossing current sensing. Consequently, the total switching period \(T\) varies with the operating point. When \(C_r\) is sufficiently large such that its voltage ripple is negligible, i.e., \(V_{C_r,\max}-V_{C_r,\min}\approx 0\), the closed-loop dynamics of the converter become similar to those of a series-capacitor buck converter~\cite{qianFastResponseVariableFrequencySeriesCapacitor2025}. Under this condition, the resonant converter can be treated as a variable frequency series-capacitor buck converter and analyzed using a switching-synchronized sampled-state space approach~\cite{cuiFastResponseVariableFrequency2023b}, \cite{qianFastResponseVariableFrequencySeriesCapacitor2025}.
However, when \(v_{C_r}\) cannot be approximated as a DC quantity, the converter modeling becomes significantly more complicated, which may obscure the insight required for linear controller design. Therefore, in this work, the parameters of the PI controller are tuned using an RL algorithm~\cite{ghamariDesignAdaptiveRobust2024,hajihosseiniDCDCPower2020}. Specifically, the converter model is reconstructed in simulation using \eqref{eq:sol1}-\eqref{eq:sol2}, and a Proximal Policy Optimization (PPO)-based algorithm is employed to optimize the PI gains. The reward function is designed to account for voltage tracking error, settling time under load disturbances, and the behavior of the control variable \(T_1\). In particular, penalties are introduced when \(T_1\) reaches its saturation limits or exhibits oscillatory behavior, thereby discouraging unstable operation and promoting desirable closed-loop dynamics. Using the optimized controller gains, the simulated time-domain waveform for 48-V-to-6-V operation at \(I_{\mathrm{load}}=20~\mathrm{A}\) under a 10\% load disturbance is shown in Fig.~\ref{fig:trans}, where a settling time of six switching cycles is achieved.

\section{Experimental Results}

\begin{figure}[t]
    \centering
    \includegraphics[width=0.96\linewidth]{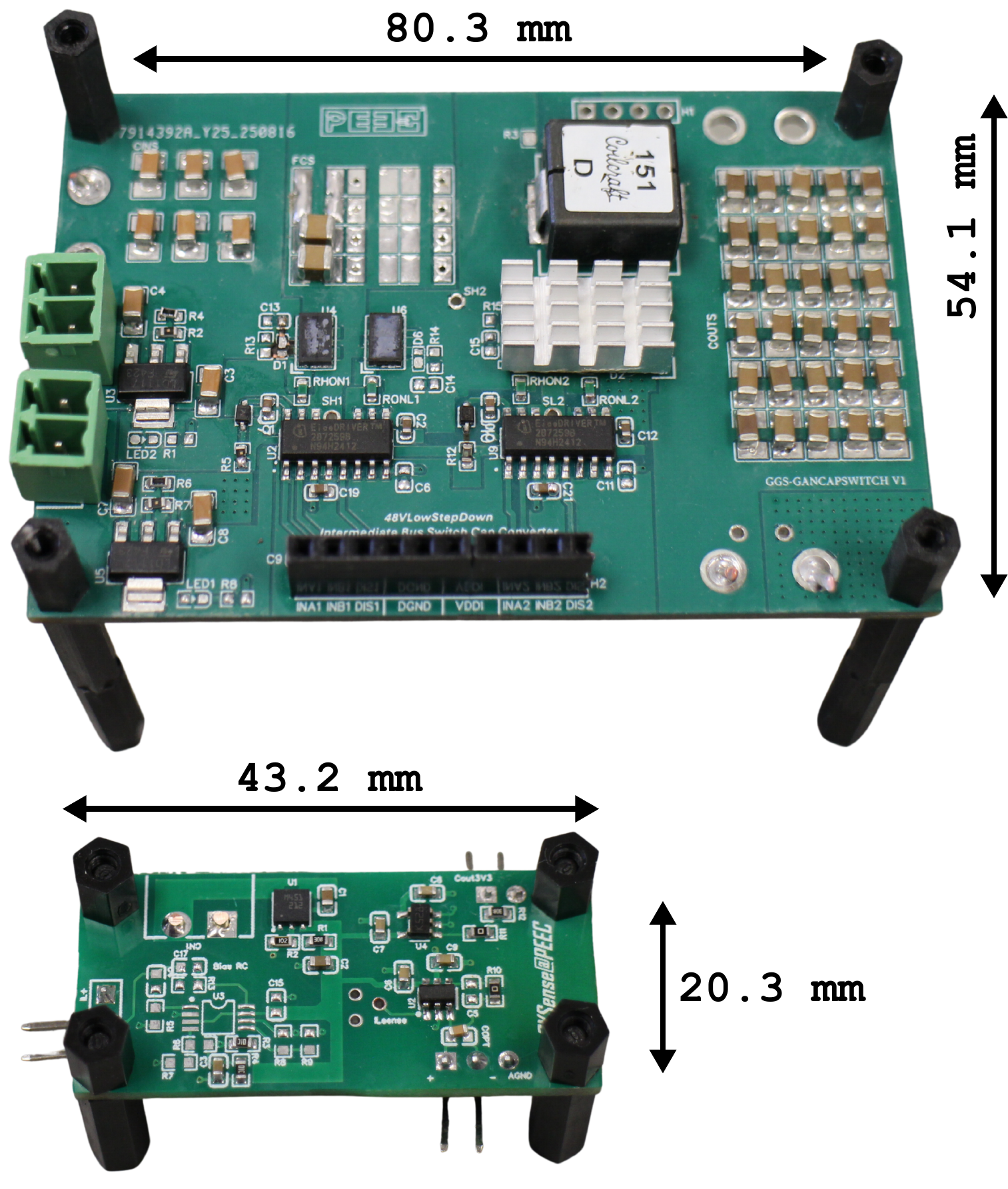}
    \caption{Photograph of the hardware prototype, including the HSCC and the signal-sensing board integrating the current-sense amplifier and comparator.}
    \label{fig:hardware}
\end{figure}

An HSCC prototype was built and tested to verify the proposed ZCS and ZVS modulation schemes. For safe operation, the experimental validation was conducted at a reduced input voltage of 24~V, while validation at the full 48-V input is left for future work. Photographs of the hardware prototypes are shown in Fig.~\ref{fig:hardware}, along with a detachable current-sensing board, and the corresponding specifications are listed in Table~\ref{tab:specs}. The current-sensing module uses DCR sensing~\cite{Tian2020AApplications}, which eliminates the need for a dedicated shunt resistor and thus reduces conduction loss and improves efficiency. The sensed current signal is conditioned and processed by a current-sense amplifier, followed by a comparator~\cite{qianControlOrientedModelingComparator2026}, which reliably detects the zero-current instants and generates the control logic required for ZCS soft-switching transitions. It is worth noting that the overall converter weight is 22~g, which is significantly lower than that of comparable transformer-based designs.

\begin{table}[t]
\centering
\caption{Prototype Converter Specifications}
\label{tab:specs}
\renewcommand{\arraystretch}{1.2}
\small
\resizebox{\linewidth}{!}{%
\begin{tabular}{ll}
    \toprule
    \textbf{Parameter} & \textbf{Value} \\
    \midrule
    Input voltage, \(V_{\mathrm{in}}\) & \(24~\mathrm{V}\) \\
    Output voltage, \(V_o\) & \(5\text{--}10~\mathrm{V}\) \\
    Output current, \(I_o\) & \(4\text{--}12~\mathrm{A}\) \\
    Switching frequency, \(f_{\mathrm{sw}}\) & \(110\text{--}122~\mathrm{kHz}\) \\
    GaN FETs & 4\(\times\) IGC033S101 \(\left(R_{\mathrm{DS(on)}} = 2.4~\mathrm{m\Omega},\; C_{\mathrm{OSS}} = 43~\mathrm{nF}\right)\) \\
    Diodes & 4\(\times\) V5NL63-M3/I \\
    Resonant capacitor, \(C_r\) & \(20~\mu\mathrm{F}\), 2\(\times\) GRM32EC72A106KE05L \\
    Resonant inductor, \(L_r\) & \(150~\mathrm{nH}\) (Coilcraft SLC1480-151MLD, \(R_{\mathrm{dc}} = 0.18~\mathrm{m\Omega}\)) \\
    Current-sense amplifier & TI INA299A1IDBVR \\
    Comparator & ST TS3021HIYLT \\
    Total converter weight & \(22~\mathrm{g}\) \\
    \bottomrule
\end{tabular}}
\end{table}

\begin{figure}[t]
    \centering
    \includegraphics[width=0.96\linewidth]{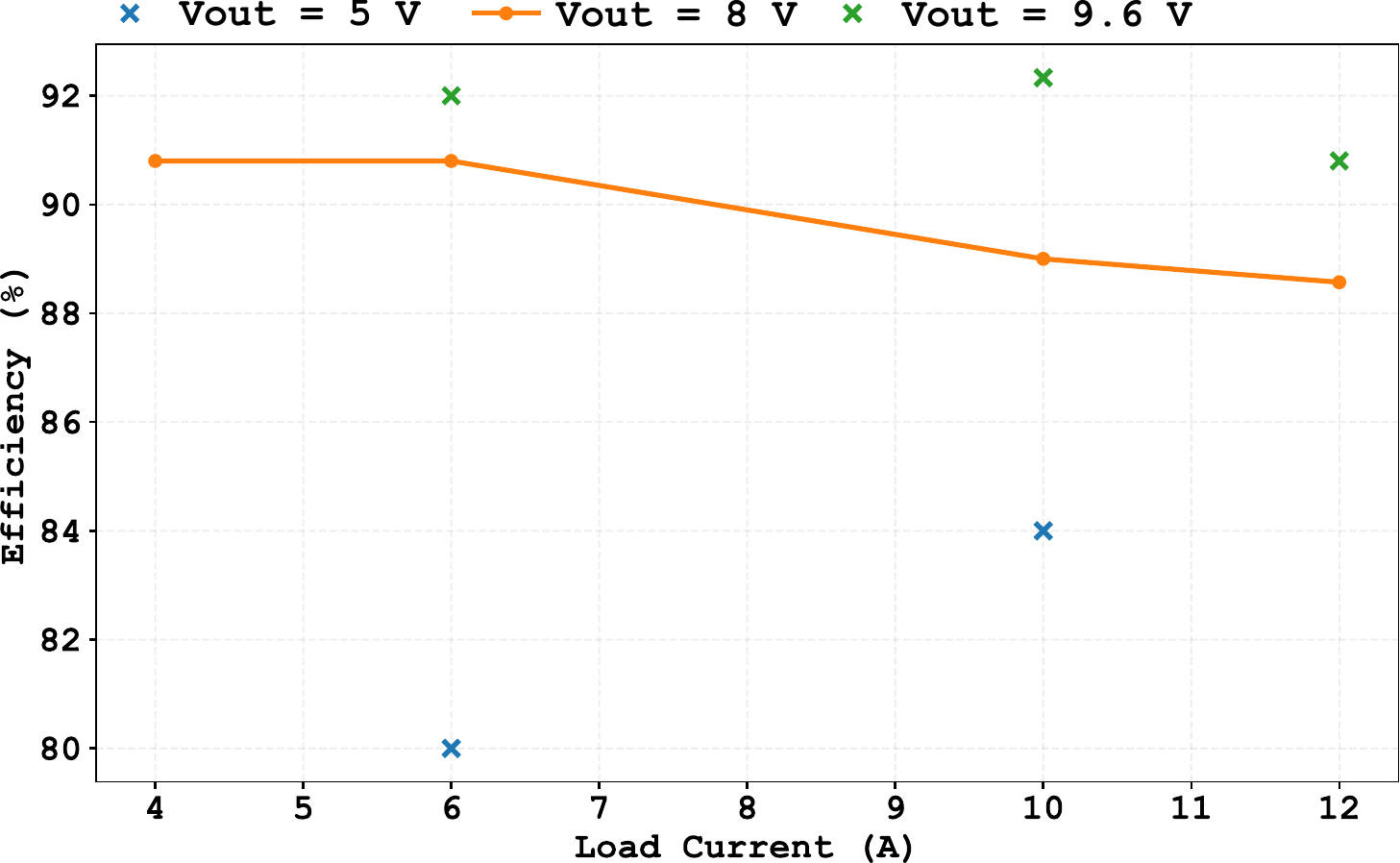}
    \caption{Measured efficiency under ZCS modulation as a function of load current at \(V_{\mathrm{in}}=24~\mathrm{V}\) for \(V_{\mathrm{out}}=5~\mathrm{V}\), \(8~\mathrm{V}\), and \(9.6~\mathrm{V}\), corresponding to voltage conversion ratios ranging from 0.2 to 0.4.}
    \label{fig:ZCS E}
\end{figure}

\begin{figure}[t]
    \centering
    \includegraphics[width=0.96\linewidth]{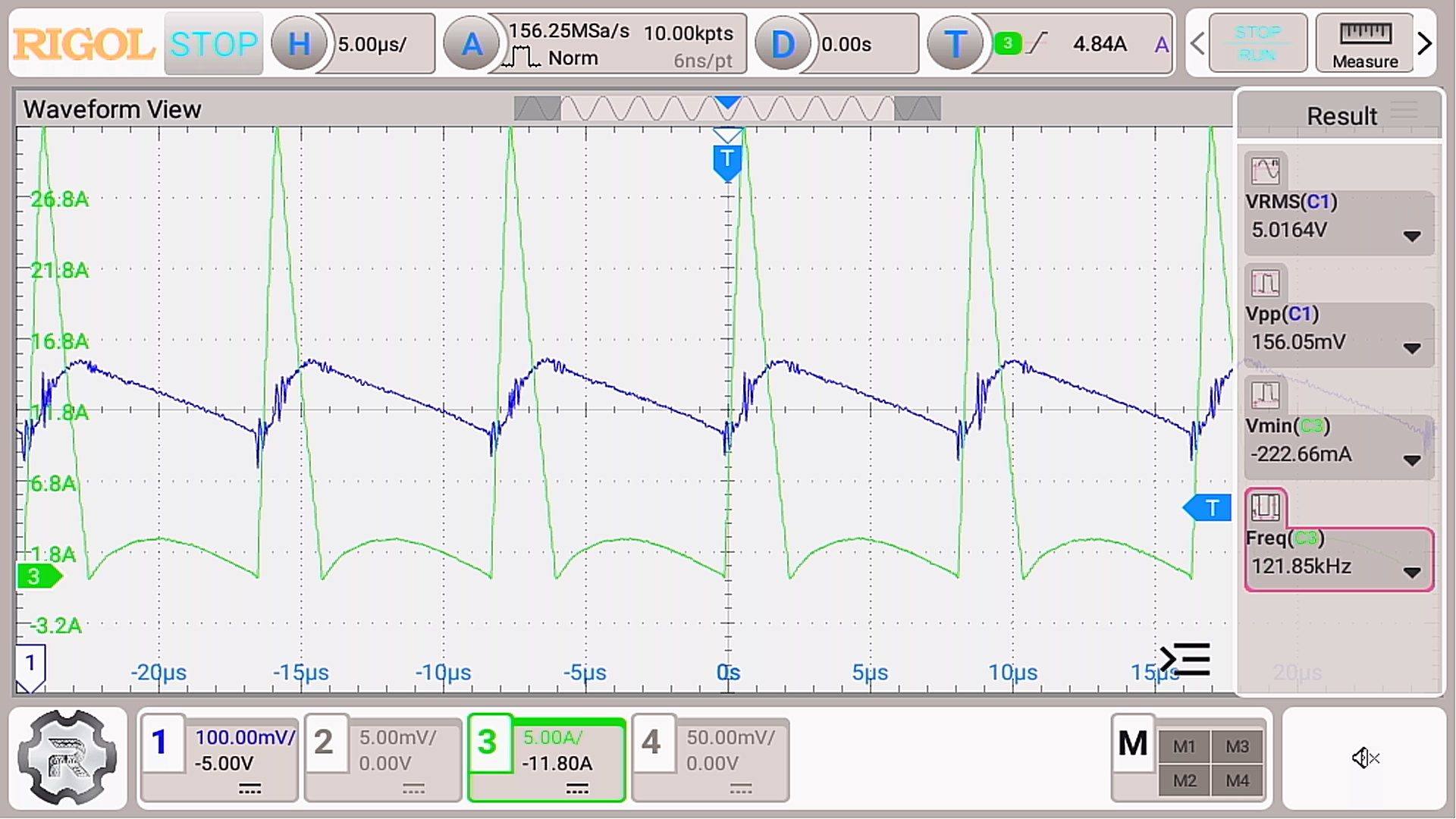}
    \caption{Measured inductorcurrent (Ch3, green) and output-voltage (Ch1, blue) waveforms under ZCS operation at \(V_{\mathrm{in}}=24~\mathrm{V}\) and \(V_{\mathrm{out}}=5~\mathrm{V}\).}
    \label{fig:ZCS WAVEFORM1}
\end{figure}

\begin{figure}[t]
    \centering
    \includegraphics[width=0.96\linewidth]{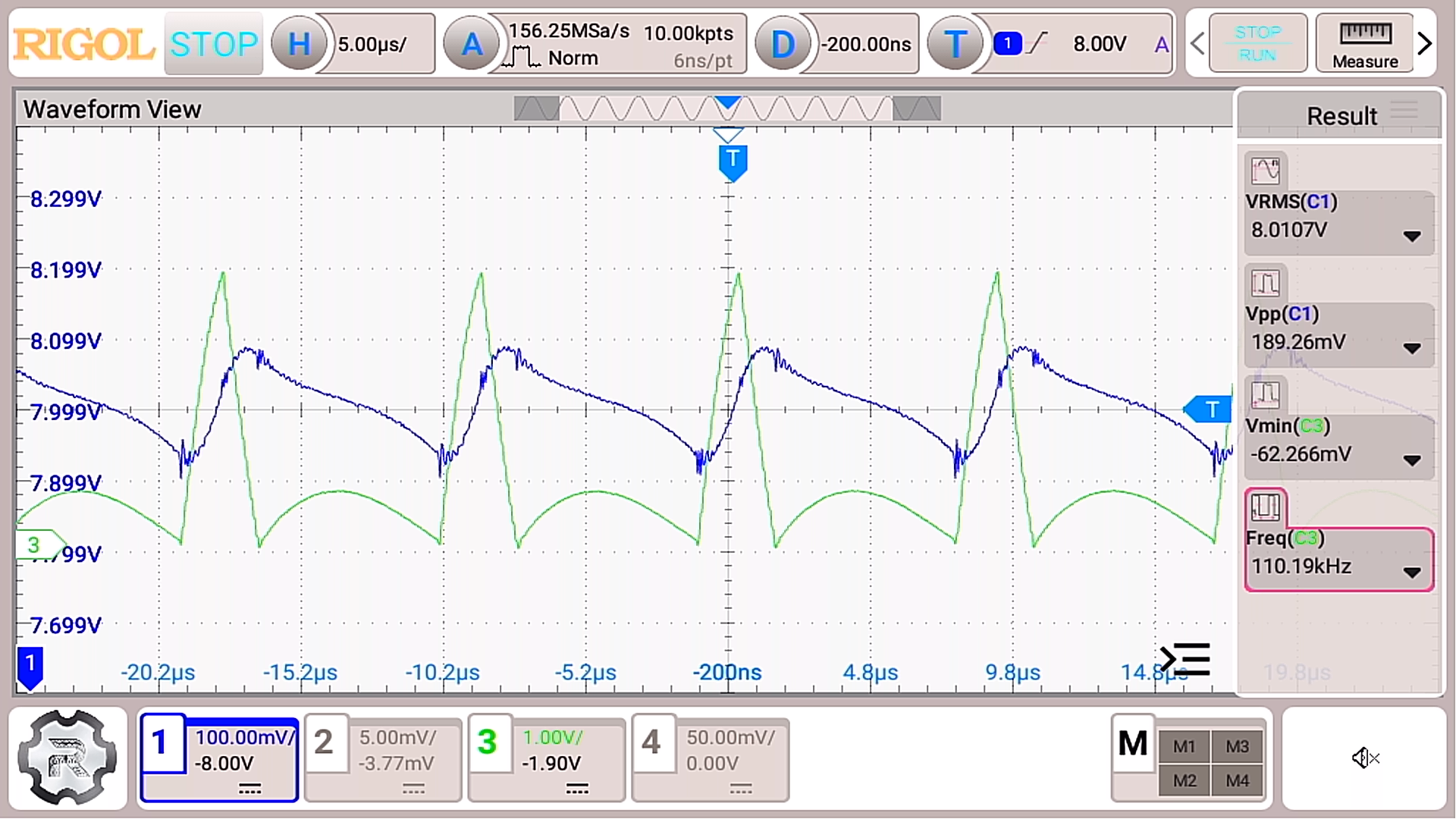}
    \caption{Measured inductor current (Ch3, green) and output-voltage (Ch1, blue) waveforms under ZCS operation at \(V_{\mathrm{in}}=24~\mathrm{V}\), \(V_{\mathrm{out}}=8~\mathrm{V}\).}
    \label{fig:ZCS WAVEFORM2}
\end{figure}

\begin{figure}[t]
    \centering
    \includegraphics[width=0.96\linewidth]{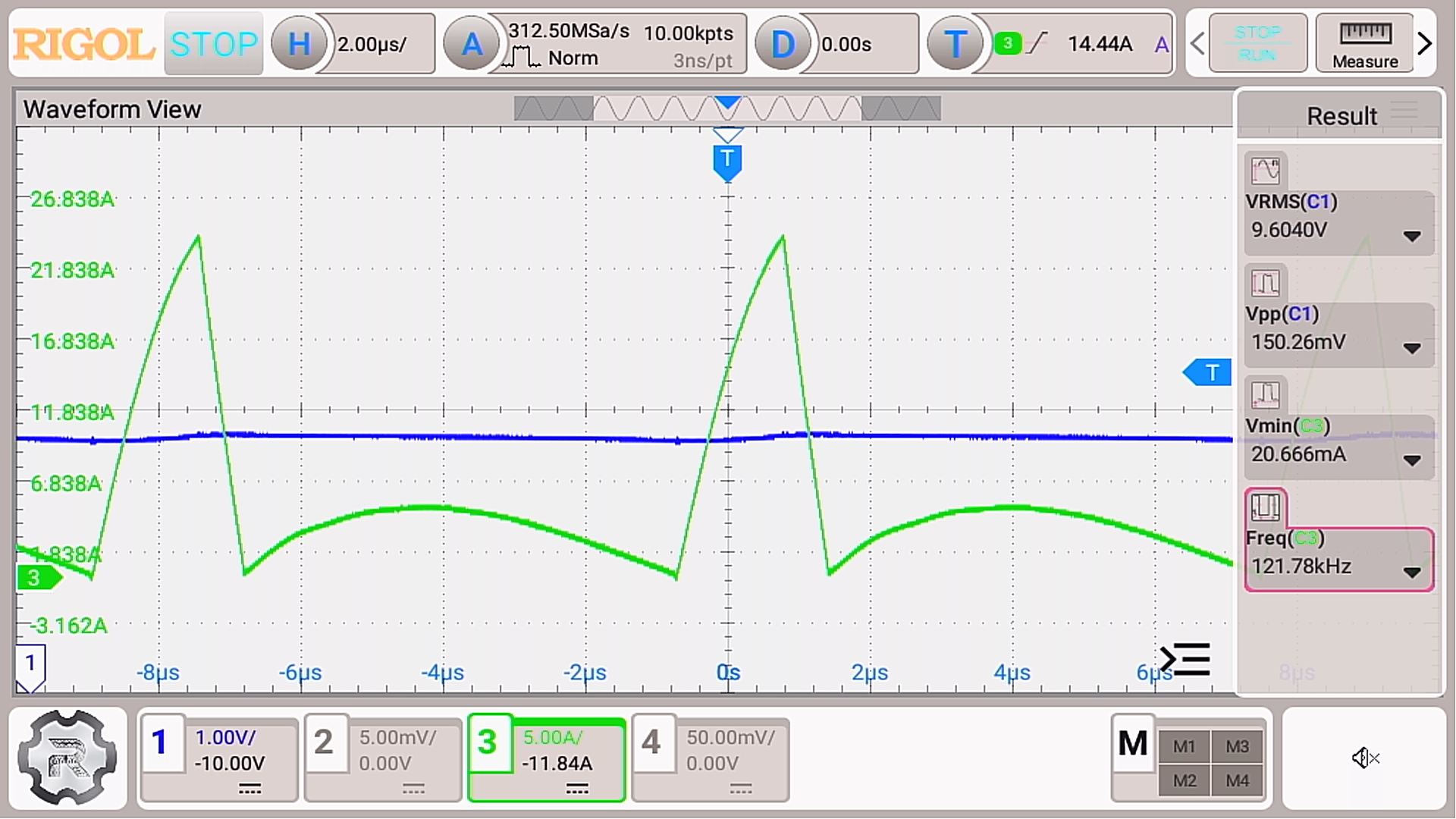}
    \caption{Measured inductor current (Ch3, green) and output-voltage (Ch1, blue) waveforms under ZCS operation at \(V_{\mathrm{in}}=24~\mathrm{V}\) and \(V_{\mathrm{out}}=9.6~\mathrm{V}\).}
    \label{fig:ZCS WAVEFORM3}
\end{figure}

For the ZCS operation, testing was carried out over a voltage conversion range of 0.2 to 0.4. The measured waveforms are presented in Fig.~\ref{fig:ZCS WAVEFORM1}-\ref{fig:ZCS WAVEFORM3}, showing the inductor current, output voltage, output voltage ripple, and switching frequency. At each voltage conversion ratio, measurements were obtained at output currents of 6~A, 10~A, and 12~A to evaluate performance across different operating points. The corresponding efficiency results are shown in Fig.~\ref{fig:ZCS E}. Efficiency above 90\% is achieved at voltage conversion ratios of 0.3 and 0.4 over the 30--100-W power range. At the lowest conversion ratio of 0.2, however, the large peak inductor current during the transition from State~1 to State~2 causes elevated conduction loss because of the non-ZCS switching of the high-side switch \(S_1\). Consequently, a noticeable reduction in efficiency is observed. This limitation could be alleviated by adopting a multiphase operation.

\begin{figure}[t]
    \centering
    \includegraphics[width=0.96\linewidth]{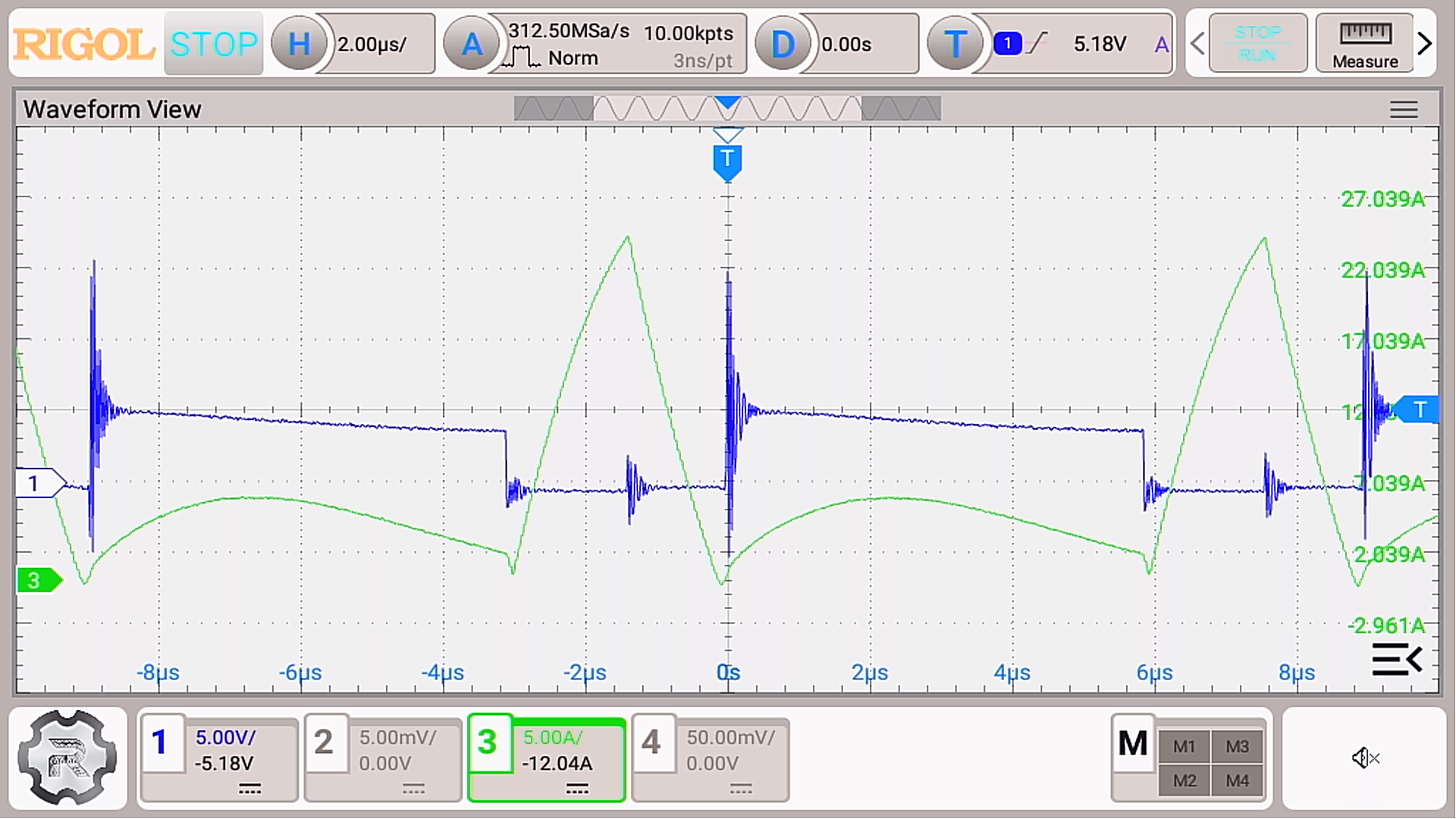} 
    \caption{Measured inductor current (Ch3, green) and switching-node voltage \(V_{\mathrm{SW}}\) (Ch1, blue) waveforms under ZVS operation.}
    \label{fig:ZVS WAVEFORM}
\end{figure}

Waveforms for the ZVS modulation scheme are presented in Fig.~\ref{fig:ZVS WAVEFORM}, showing the inductor current and the switching-node voltage \(V_{\mathrm{SW}}\). Specifically, \(V_{\mathrm{SW}}\) falls to zero before the transition to State~1, confirming successful ZVS turn-on of \(S_4\). In the present implementation, the current-sensing module was used to determine the ZCS transition, whereas the ZVS transition was manually tuned by terminating State~3 and adjusting the discharge interval \(T_{\mathrm{dis}}\) such that the inductor current remained high enough to fully discharge \(C_{\mathrm{OSS}}\). Although the efficiency improvement under ZVS operation is not investigated in detail in this paper, the proposed ZVS scheme is expected to be particularly beneficial under light-load conditions~\cite{Liang2022LightControl}, where the reduction in turn-on switching loss becomes more important for maintaining high efficiency.


\section{Conclusion}

In this work, a novel multi-mode modulation scheme for a hybrid switched-capacitor converter is proposed to extend voltage regulation beyond the conventional 2:1 conversion ratio while preserving soft-switching operation in most transitions. The proposed approach is implemented using a three-state operating scheme that supports both ZCS and ZVS transitions, maintains output voltage regulation, and enables variable step-down ratios. Experimental results from a 24 V prototype validate the proposed modulation scheme, demonstrating efficient operation over voltage conversion ratios of 0.3--0.4, with a peak efficiency exceeding 92\% at 100~W. The feasibility of ZVS operation is also confirmed, although further investigation is required to fully assess its performance over a broader range of operating conditions. Overall, the proposed modulation scheme significantly expands the practical regulation range of resonant switched-capacitor converters while maintaining high efficiency.

\bibliographystyle{ieeetr}  %
\bibliography{references}  
\end{document}